\documentclass[conference]{IEEEtran}
\IEEEoverridecommandlockouts
\usepackage{cite}
\usepackage{amsmath,amssymb,amsfonts}
\usepackage{algorithmic}
\usepackage{graphicx}
\usepackage{textcomp}
\usepackage{xcolor}
\usepackage{multirow}
\usepackage{subcaption}

\def\BibTeX{{\rm B\kern-.05em{\sc i\kern-.025em b}\kern-.08em
    T\kern-.1667em\lower.7ex\hbox{E}\kern-.125emX}}
\usepackage{array} 
\begin{document}

\title{Designing Reputation Systems for Manufacturing Data Trading Markets: A Multi-Agent Evaluation with Q-Learning and IRL-Estimated Utilities\\
}
                
\author{\IEEEauthorblockN{Kenta Yamamoto}
\IEEEauthorblockA{\textit{ Department of Systems Innovation, 
} \\
\textit{Graduate School of Engineering,}\\
The University of Tokyo \\
Tokyo, Japan\\
4828556767@g.ecc.u-tokyo.ac.jp}
\and
\IEEEauthorblockN{Teruaki Hayashi}
\IEEEauthorblockA{\textit{Department of Systems Innovation,
} \\
\textit{Graduate School of Engineering,}\\
The University of Tokyo\\ 
Tokyo, Japan\\
hayashi@sys.t.u-tokyo.ac.jp}
}

\maketitle

\begin{abstract}
Recent advances in machine learning and big data analytics have intensified the demand for high-quality cross-domain datasets and accelerated the growth of data trading across organizations. As data become increasingly recognized as an economic asset, data marketplaces have emerged as a key infrastructure for data-driven innovation. However, unlike mature product or service markets, data-trading environments remain nascent and suffer from pronounced information asymmetry. Buyers cannot verify the content or quality before purchasing data, making trust and quality assurance central challenges.
To address these issues, this study develops a multi-agent data-market simulator that models participant behavior and evaluates the institutional mechanisms for trust formation. Focusing on the manufacturing sector, where initiatives such as GAIA-X and Catena-X are advancing, the simulator integrates reinforcement learning (RL) for adaptive agent behavior and inverse reinforcement learning (IRL) to estimate utility functions from empirical behavioral data. Using the simulator, we examine the market-level effects of five representative reputation systems—Time-decay, Bayesian-beta, PageRank, PowerTrust, and PeerTrust—and found that PeerTrust achieved the strongest alignment between data price and quality, while preventing monopolistic dominance. Building on these results, we develop a hybrid reputation mechanism that integrates the strengths of existing systems to achieve improved price–quality consistency and overall market stability.
This study extends simulation-based data-market analysis by incorporating trust and reputation as endogenous mechanisms and offering methodological and institutional insights into the design of reliable and efficient data ecosystems.

\end{abstract}

\renewcommand{\IEEEkeywordsname}{Keywords} 

\section{Introduction}
Recent advances in machine learning and big data analytics have increased the demand for diverse high-quality datasets across industries. As sensing, networking, and distributed computing technologies mature, massive amounts of heterogeneous data are generated and shared in real time. To leverage these data for innovation, organizations increasingly participate in data trading beyond institutional boundaries, giving rise to data marketplaces that function as the foundational infrastructure for a data-driven economy [1–6].

To address these challenges, a simulator that can model the key elements of data markets and reproduce participant behavior under diverse rules and regulations is required [7-9]. Accordingly, we developed a data-market simulator that integrates multi-agent reinforcement learning (RL) and inverse reinforcement learning (IRL) to model participant behavior and evaluate the institutional mechanisms for trust formation. Focusing on the manufacturing domain, where large-scale data-sharing initiatives, such as GAIA-X and Catena-X, are advancing, we designed a simulation framework that reflects real-world decision dynamics in data trading. In the proposed approach, IRL is used to estimate the utility of participants based on behavioral traces on a production data platform, whereas RL enables agents to form and adapt their strategies over time.

Using this simulator, we examined the market-level impacts of five representative reputation systems—Time-decay, Bayesian-beta, PageRank, PowerTrust, and PeerTrust—which are widely studied in the areas of trust management and distributed systems [10–14]. The results demonstrated that PeerTrust yielded the highest alignment between data prices and quality. Building on these findings, we developed a novel hybrid reputation system that integrates the strengths of the existing models and achieves improved market consistency.

This study makes three contributions.
\begin{enumerate}
    \item We developed a simulation-based experimental platform to analyze institutional mechanisms in data markets using ML-based behavioral modeling.
    \item We conducted a comparative analysis of reputation systems under large-scale data trading scenarios.
    \item We proposed a new hybrid reputation mechanism to enhance trust and market efficiency.
\end{enumerate}
This study bridged the domains of machine learning, multi-agent simulation, and big data ecosystem design, offering a computational framework to understand how reputation and trust can sustain a scalable and reliable data marketplace.

\section{Related Works and Our Approach}
\subsection{Data Trading Markets and Their Challenges}
The emergence of data trading markets, which are also referred to as data marketplaces, has been widely discussed as a new form of the digital economy [1–6]. These markets enable cross-domain data exchange and the economic use of data assets. Studies have proposed technologies that facilitate such exchanges, including pricing mechanisms [15,16] and data-protection frameworks [6]. However, despite growing academic and industrial interest, data markets remain in their infancy, with limited observable phenomena. Institutional structures and market rules are still evolving, and a comprehensive understanding of the optimal design principles is lacking. Moreover, information asymmetry poses a major challenge; it is difficult to evaluate data quality or verify content before purchase. Consequently, ensuring data reliability and participant trust remains a crucial, yet unresolved issue in data market design.

\subsection{Trust and Reputation in Online and Data Markets}
To address similar challenges in product and service markets, online platforms have long employed reputation systems to ensure quality and foster trust among market players. Since the early days of e-commerce, platforms such as eBay and Amazon have demonstrated that trustworthy sellers receive higher ratings and that better reputations correlate with increased transaction prices [17,18].

Subsequent research has examined the mechanisms by which reputation influences purchasing decisions and explored review bias, trust propagation, and rating fairness [19–22]. Although these systems encourage participation, they also exhibit inflation and bias in their ratings. To mitigate these effects, algorithms that weigh reviewers based on credibility or past behavior have been proposed, including those used on Q\&A platforms such as Stack Overflow [23].

More recent studies have focused on graph-based trust propagation, including EigenTrust [12], which applies PageRank principles and probabilistic models such as Beta Reputation [11]. Parallel efforts have addressed fake review detection and manipulation using linguistic, behavioral, and social graph features to enhance robustness [24,25]. In the context of data-trading markets, these issues become even more critical: data quality cannot be verified before data purchase and analysis, and transactions often depend on the perceived trustworthiness of providers [18]. 

\subsection{Simulation Approaches for Market and Institutional Design}
Simulation-based analysis has emerged as a principal method for studying market-level dynamics and evaluating institutional mechanisms such as reputation systems. Multi-agent simulations enable controlled experimentation with market rules, revealing macro-level effects from micro-level behaviors [26]. Previous studies have demonstrated the potential of such approaches to capture phenomena that are otherwise unobservable, thereby supporting the design and validation of market institutions.

Fernandez et al. [27] emphasized the importance of adopting simulation-based methods in data-market research. For instance, a previous study constructed a network representation of a market in which traded datasets were modeled as nodes consisting of variable sets, and scenario analyses of data-purchasing behaviors were performed [28]. Another study proposed a simulation framework in which buyer agents perform random walks on a data network to reproduce the relationships between datasets, thereby assessing the impact of different purchasing strategies [29]. Research is also being conducted on agent-based data-market simulations that enable data replication and resale [30].

However, these studies generally overlook the role of trust and reputation in shaping market outcomes. In reality, particularly in markets characterized by information asymmetry, perceptions of the reliability of data providers and datasets have a decisive impact on trading decisions. The absence of reputation mechanisms in existing simulations limits their ability to capture the process through which trust is built, propagated, and reflected in price formation.

\subsection{Summary and Our Approach}
To address this gap, this study incorporated reputation systems as endogenous market mechanisms within a multi-agent simulation framework. By integrating RL for adaptive decision making and IRL for empirically grounded utility estimation, our approach enables agents to learn and evolve strategies based on market feedback and reputational information. This design allows for a more realistic reproduction of trust-driven behaviors in data trading and provides a computational environment for testing institutional designs that promote both market efficiency and reliability.

Consequently, this study contributes to the literature by extending simulation-based analyses of data markets from a purely economic or structural perspective to one that explicitly models trust formation and reputation dynamics, a critical yet previously underexplored dimension of data-exchange ecosystems.

\section{Models of Data Marketplace}
\subsection{Our Market Model}\label{AA}
This study constructed a data-trading market model primarily focused on the manufacturing sector, where the need for cross-organizational data sharing is rapidly increasing. The market was comprised of four main entities: data providers, buyers, marketplaces, and platform operators. Simulations assumed that these participants interacted within an open marketplace, where transactions were freely conducted among diverse organizations.

Because manufacturing data emphasize not only technical accuracy but also organizational and regulatory aspects, four quality dimensions—accuracy, coverage, freshness, and compliance—are defined as key indicators of data value [31,32]. These characteristics reflect the distinct requirements of manufacturing environments, where traceability, reliability, and interoperability are critical for data utilization.

Each provider adopts one of four strategic orientations—price, trust, quality, or standard-oriented—which represent different business priorities within the market. Buyers employ three strategies—price, quality, and trust—to select the datasets that best satisfy their utility functions.

Simulations were conducted under the assumption that the market was open rather than closed. This open structure enabled us to evaluate  how diverse agents with various strategies interacted, competed, and adapted to different market rules and reputation mechanisms.

\subsection{Trading Model}
The trading model in this study formalized the interactions between four types of market participants: providers, buyers, the marketplace, and the platform operator, as illustrated in Fig. 1. The marketplace acted as an intermediary that hosted transactions between providers and buyers, whereas the platform operator oversaw the governance functions of the market, such as data quality management, compliance verification, and the operation of the reputation system.

\begin{figure}[h]
    \centering
    \includegraphics[width=1.0\linewidth]{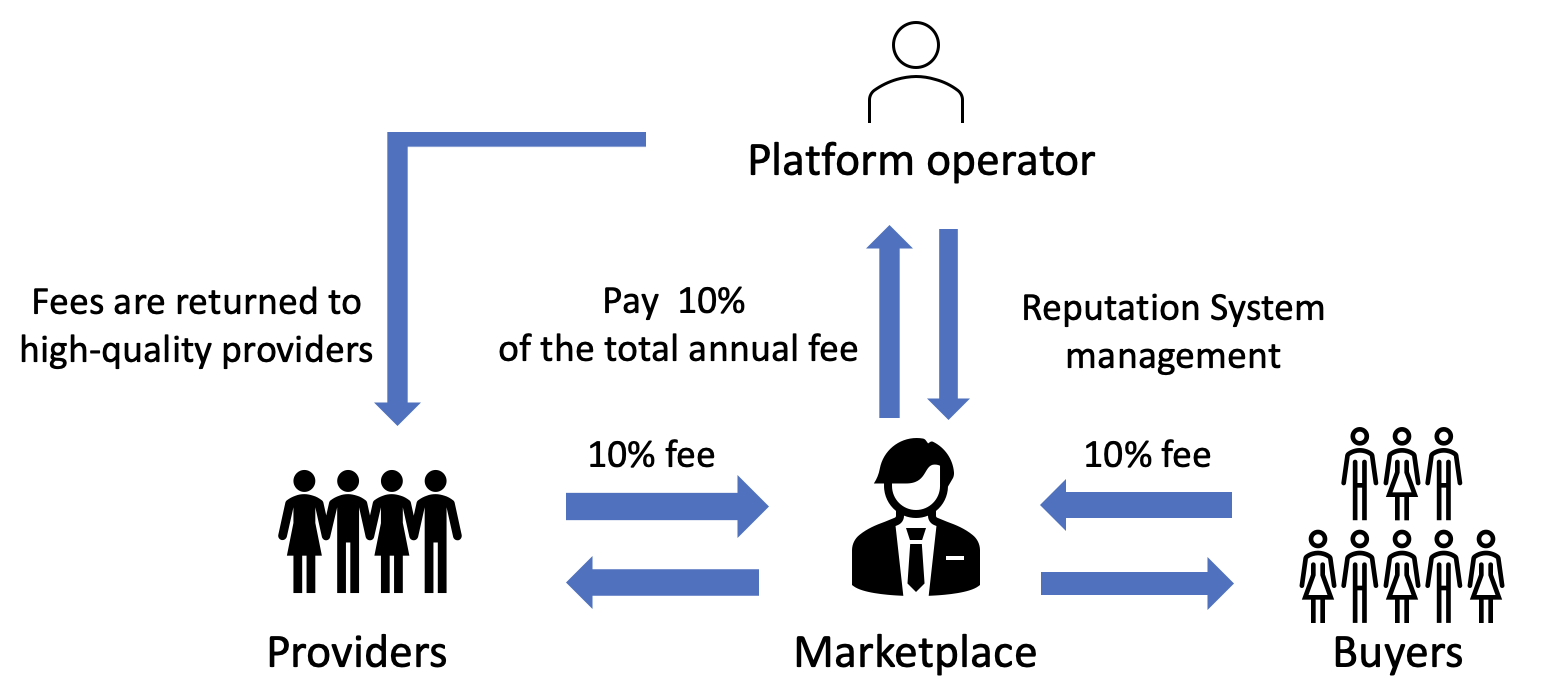}
    \caption{Trading Model of the Data Market. }
    \label{fig:placeholder}
\end{figure}

The market configuration was parameterized using data from the Teikoku Databank\footnote{https://www.tdb-en.jp/index.html}, a Japanese think tank, and the numbers of providers and buyers were set to $2,000$ each, based on the scale of Tier 1 firms in the major manufacturing groups. 

During a transaction, the provider and buyer both pay a $10\%$ transaction fee to the marketplace. The marketplace, in turn, transfers $10\%$ of its annual revenue to the platform operator, who administers the compliance and quality assurance functions. To incentivize responsible behavior, providers offering highly compliant and reliable data receive a full refund from the platform operator.

The fee and governance flow model is an institutional framework that balances market efficiency and trust assurance. It also enables the simulator to evaluate how financial incentives, compliance mechanisms, and reputation-based interactions collectively influence the data market stability and quality evolution.

\subsection{Data Quality}
This study assumed a data-trading market that primarily dealt with manufacturing data, which are characterized by strict requirements for accuracy and compliance. Accordingly, data quality was defined along four key dimensions: accuracy, coverage, freshness, and compliance, which together represented both the technical and governance aspects of data reliability [31,32]. For each provider, a data quality vector was generated to represent the quality attributes of the datasets offered in the market, as shown in Eq.(1). 


\begin{equation}
\mathbf{q} =
\begin{pmatrix}
\text{accuracy} & \text{freshness} \\
\text{coverage} & \text{compliance}
\end{pmatrix}
\in [0,1]^4
\end{equation}

\noindent These quality scores were randomly sampled from normal distributions where the mean and variance differed according to the strategy of the provider.

This approach allowed the simulation to reproduce the heterogeneity in data quality across providers, reflecting real-world manufacturing environments in which datasets vary widely in terms of precision, scope, update frequency, and adherence to standards. 

\subsection{Data Pricing and Cost}

In this study, data prices were modeled as stochastic variables generated according to the strategic behavior of each provider, similar to the way the data quality was treated. Specifically, prices were sampled from normal distributions, whose mean values were adjusted by strategy-specific multipliers that represented different market strategies. The production cost associated with providing the dataset was calculated using a quadratic cost function [33], as shown in Eq.(2):

\begin{equation}
C(q)=
b\bigl(
    c
    + d\, q_{\text{acc}}^{2}
    + e\, q_{\text{comp}}^{2}
    + f\, q_{\text{fresh}}^{2}
    + g\, q_{\text{cov}}^{2}
\bigr),
\end{equation}

\noindent where $C$ denotes the cost and $q_{\text{acc}}$, $q_{\text{comp}}$, $q_{\text{fresh}}$, and $q_{\text{cov}}$ represent the quality attributes of accuracy, compliance, freshness, and coverage, respectively, which each had a weight in the range of $b$ through $g$. In this formulation, accuracy and compliance were assigned higher weights than freshness and coverage, reflecting their greater importance in manufacturing data transactions, where precision and regulatory adherence strongly influence the perceived value.

\subsection{Market Strategies and RL}
In the proposed data-trading market, the participants were divided into two primary agent types: providers and buyers, which were each equipped with distinct strategic options. Providers could adopt one of four strategies: trust oriented, price oriented, quality oriented, or standard. Buyers could choose one of three strategies: trust, price, or quality. These strategies determined the behavioral priorities of the agents.

\begin{itemize}
\item The trust-oriented strategy emphasized compliance and accuracy but maintained relatively low prices to gain the confidence of buyers.
\item The price-oriented strategy emphasized price competitiveness, typically reducing quality to achieve cost efficiency.
\item The quality-oriented strategy maintained a high data quality and slightly higher prices, representing fair-value transactions.
\item The standard strategy served as a baseline, averaging the parameters of the other strategies.
\end{itemize}

To simulate adaptive decision making, both providers and buyers updated their strategies through RL. Specifically, a $\varepsilon$-greedy Q-learning algorithm was implemented to allow agents to balance exploration and exploitation while learning optimal actions. This approach overcame the rigidity of threshold-based rules and enabled agents to respond dynamically to changing market conditions, which is a technique widely applied in financial and multiagent simulations [34,35].

The Q-value of the state–action pair $(s, a)$ of an agent was updated according to the Bellman equation:
\begin{equation}
Q(s,a) \leftarrow Q(s,a) + \alpha \bigl[ r + \beta \cdot \max_{a'} Q'(s',a') - Q(s,a) \bigr],
\end{equation}

\noindent where $\alpha$ is the learning rate, $\beta$ is the discount factor, $r$ is the agent's reward, and the exploration rate is $0.995$. The provider and buyer rewards are expressed by Eqs.(4) and (5) in Subsection F, respectively.

For providers, the state was defined by the cumulative profit and recent reputation, whereas the reward function reflected the profit adjusted for transaction fees, fee refunds, and buyer evaluations. Buyer states were represented by their recent transaction success rates and average utility, enabling them to adaptively revise their purchasing strategies based on their accumulated experience.

\subsection{Utility Functions and IRL}
In the RL framework described in the previous section, the reward function of the provider combined the economic performance and trust-related feedback. As defined in Eq.(4), reward $r$ reflects not only the profit of the provider, but also the transaction costs, fee refunds, and utility of the buyer:
\begin{equation}
\begin{split}
r = \max(0,\, p - C - F)\,\cdot (1 + F_r)\\
\quad {}+ h\cdot\,U + k\cdot\,\max(0,\, U - R_{\text{before}}),
\end{split}
\end{equation}

\noindent where $p$ denotes the data price, $C$ is the cost, $F$ is the transaction fee, $F_r$ is the fee refund, $U$ is the utility of the buyer, and $R_{\text{before}}$ is the reputation of the provider generated immediately before a transaction in each reputation system described in subsection G. In addition, $h$ and $k$ represent weights. This formulation allowed providers to balance profit maximization and reputation management, ensuring that actions that promoted higher perceived quality or trust were rewarded over time.

To operationalize $U$, the utility function of the buyer was defined as a weighted combination of three core components: the data quality, reputation, and price:
\begin{equation}
U = l \cdot q + o \cdot R_{\text{before}} + u \cdot p,
\end{equation}

\noindent where $q$ denotes the data quality, $R_{\text{before}}$ denotes the reputation of the provider, as in Eq.(4), and $p$ denotes the data price. The coefficients $l$, $o$, and $u$ represent the relative importance values of these factors and were estimated empirically rather than given arbitrary values. The value of this utility function was directly used as a reward for the RL agent.

To obtain realistic weights, we applied IRL to behavioral data from Kaggle, the world’s largest data competition platform, which provides a proxy for buyer decision-making in data markets. Kaggle has various features such as kernels, which are notebooks with an execution environment that runs entirely in the browser, along with forums for each competition and dataset, where participants can exchange opinions and engage in a wide range of activities. From the Meta-Kaggle dataset (May 2024 version), we extracted $500$ users who created relatively few datasets but voted frequently, which is a behavior akin to that of buyers in the data-trading market. Using the Maximum Causal Entropy IRL (Max-Causal-Ent-IRL) algorithm [36,37], we estimated the latent reward parameters that best explained the observed user behavior, while accommodating non-optimal or stochastic actions. Using seven action types for these users (forum creation, kernel creation, submissions, dataset voting, kernel voting, forum voting, and dataset creation), we estimated the weight of each action using IRL. We defined each action $a$ as one of these seven behaviors and state $s$ as the cumulative action count. Then, using feature vector $\phi$ constructed from $(s,a)$ and the reward weights, $\theta$, we specified linear reward $r_{\theta}$, as shown in Eq.(6).

\begin{equation}
r_{\theta}(s,a) = \theta^{\top} \phi(s,a)
\end{equation}

\noindent The soft Bellman equations were applied to estimate the optimal policy.
\begin{equation}
Q_{\theta}(s,a) \leftarrow r_{\theta}(s,a)
  + \gamma \,\cdot\mathbb{E}_{s' \sim P(\,\cdot \mid s,a)}\!\bigl[ V_{\theta}(s') \bigr]
\end{equation}

\begin{equation}
V_{\theta}(s) = \log \sum_{a \in A} \exp(Q_{\theta}(s,a))
\end{equation}

\begin{equation}
\pi_{\theta}(a \mid s) = \exp(Q_{\theta}(s,a) - V_{\theta}(s))
\end{equation}

\noindent Here, $P$ denotes the probability of a transition to the next state, $s'$, when taking action $a$ in state $s$, and $\mathbb{E}$ denotes the expectation with respect to this transition. Using these definitions, we set the action value function, $Q_{\theta}(s,a)$, in Eq.(7) with discount rate $\gamma$  and soft value function $V_{\theta}$ in Eq.(8) to define the soft Bellman equations. We also specify soft policy $\pi_{\theta}$ in Eq.(9) for the computation.

Using these equations, we computed empirical feature expectation $\mu_{E}$ at time $t$ for the Kaggle data, along with model-side feature expectation $\mu_{\theta}$, as defined in Eq.(10), using discount rate $\delta$. To prevent overfitting, we introduced regularization term $\epsilon$. We then determined log-likelihood $\mathcal{L}$ based on the difference between these feature expectations, and the final weights were obtained by updating the parameters according to gradient $\nabla$ with respect to $\theta$, as shown in Eq.(11). We calculated the feature expectations for the Kaggle data and model-side feature expectations using Eqs.(10) and (11), respectively. We then determined the weights by setting the gradient to the difference between these feature expectations and updating the weights accordingly.

\begin{equation}
\mu = \mathbb{E}\!\left[ \sum_{t=0}^{\infty} \delta^{t}\,\phi(s_t,a_t) \right]
\end{equation}

\begin{equation}
\nabla_{\theta}\mathcal{L}(\theta) = \mu_{E} - \mu_{\theta} - \epsilon\,\theta
\end{equation}

\noindent This yielded the results listed in table I. Normalizing these values produced the final action-specific weights. Because Kaggle is fundamentally a data competition platform, its logs do not include actions related to purchasing data. Therefore, we treated dataset votes as proxies for dataset purchases. Discussions on data and models occurred in kernels and forums, and we interpreted these actions as indicators of the emphasis placed by participants on the quality of the data or model. Therefore, we assigned weights to actions such as kernel and forum creation in proportion to their perceived quality values. Likewise, each voting action could be interpreted as indicating the extent to which participants gave attention to the evaluations of others. Accordingly, we treated the three voting actions as reputation-related weights. The averages were used when multiple weights were available for a given construct. 

\begin{table}[h]
\caption{Output Values and Normalized Values}
\label{tab:outputs}
\centering
\begin{tabular}{l|c|c} 
\hline
\textbf{action} & \textbf{Output values} & \textbf{Normalized values} \\
\hline
Dataset creation & -0.8687 & 0.0 \\
Kernel creation  & -0.4816 & 0.2342 \\
submission       &  0.7844 & 1.0 \\
Forum post       & -0.2121 & 0.3972 \\
Dataset vote     & -0.1169 & 0.4548 \\
Forum vote       &  0.1932 & 0.6424 \\
Kernel vote      &  0.7016 & 0.9499 \\
\hline
\end{tabular}
\end{table}

\subsection{Reputation Systems}
The datasets in data markets that handle manufacturing data exhibit diverse quality characteristics. Because data-generation technologies advance periodically through innovation, the value of data changes continuously over time. Accordingly, a reputational system capable of accommodating temporal and multidimensional qualities is required. Guided by these considerations, this study selected five existing reputation systems and incorporated them into a simulation, as described in detail below. In our model, the review score of each buyer was constrained to the interval $[0, 1]$, and the parameters were calibrated such that the score magnitude aligned with the utility. Upon a successful purchase, each buyer generated a review, $R$, according to Eq.(12) using weights $v$ and $y$. These review scores were then aggregated for each provider by the five reputation systems described below to produce the final reputation scores of the provider $(R_1,\ldots,R_5)$.

\begin{equation}
R = \min(1,\max(0, v + y\cdot U))
\end{equation}

\subsubsection{Time-Decay}
Time-decay is a reputation system that calculates review values by computing a weighted average, where older evaluations receive less weight and recent evaluations are given greater weight [16]. When $w_{{1i}}$ ($i=0,1...$) is the Time-decay weight and $R_{{i}}$ is the review, the reputation score $R_{{1}}$ at time $t$ is defined by Eq.(13). Half-life value $\mathrm{half}_{\mathrm{life}}$, which was the time until the weight reached half its original value, was set to $25$. This method used Eq.(14) to determine weights, which were then used to determine the reputation score. This method is effective when quality changes over time, prevents older reputation values from indefinitely having excessive influence, and provides resilience against review manipulation. 

\begin{equation}
R_1(t) = \frac{\sum_{i} w_{{1i}} R_i}{\sum_{i} w_{{1i}}}
\end{equation}
 
\begin{equation}
w_{{1i}} \propto \exp\!\bigl(-\lambda (t - t_i)\bigr), \qquad 
\lambda = \frac{\ln 2}{\mathrm{half}_{\mathrm{life}}}
\end{equation}
 
\subsubsection{Bayesian-beta}
Bayesian-beta is a Bayesian reputation model [17]. Because simply averaging reviews can overtrust users with few observations, it statistically accounts for sample-size uncertainty. The beta distribution in Eq.(15) is a probability model of \([0,1]\) that naturally accommodates binary (and bounded) ratings. The distribution mean (\(\text{mean}\)) and a confidence weight reflecting the amount of evidence (\(\text{conf}\)) are defined in Eq.(16). Given review value \(R_{i}\) ($i=0,1...$) computed from buyer utility \(U\), we update the \(\beta\)-distribution hyperparameters by adding \(R_{i}\) to \(A\) and \(1 - R_{i}\) to \(D\). Using these updates, final reputation score \(R_2\) is computed using Eq.(17). This approach mitigates overconfidence or undervaluation until sufficient data accumulate.

\begin{equation}
f(x \mid A,D)=\frac{1}{B(A,D)}\, x^{A-1}(1-x)^{D-1},\quad 0<x<1
\end{equation}

\begin{equation}
\mathrm{mean}=\frac{A}{A+D},\qquad
\mathrm{conf}=\frac{A+D}{A+D+2}\in(0,1)
\end{equation}

\begin{equation}
R_2=0.5\bigl(\mathrm{mean}+\mathrm{conf}\cdot\mathrm{mean}+0.5(1-\mathrm{conf})\bigr)
\end{equation}

\subsubsection{ PageRank}
PageRank was originally developed as a Google search engine [18]. It treats the link structure between webpages as a Markov chain and assigns an importance score to each page based on the number of inbound links and importance of the linking pages. While Time-decay and Bayesian-beta ignore the influence of the reviewer (i.e., the person who provided the review) and thus may fail to fully reflect network effects, PageRank explicitly considers this influence. In this study, we created a directed graph when a transaction occurred, where the provider was denoted by $\mathrm{PR}(i)$ and the buyer by $\mathrm{PR}(j)$. Let $N$ denote the number of providers and buyers, $m(i)$ be the set of buyers linked to $\mathrm{PR}(i)$, $L(i)$ be the number of transactions linking buyer $j$ to provider $i$, $\zeta$ be the transition probability, and $w_{3ij}$ be the weight based on the number of transactions. The PageRank value calculated using this formula was normalized to generate the final review value,  \(R_3\).

\begin{equation}
\mathrm{PR}(i)
= \frac{1- \zeta}{N}
  +  \zeta\cdot\sum_{j \in m(i)} \frac{\mathrm{PR}(j)}{L(j)}\, w_{3ij}
\end{equation}

\subsubsection{PowerTrust}
PowerTrust aggregates weights based on reviewer influence, similar to PageRank, but with the additional feature of strongly emphasizing trusted hub reviewers [19]. Simply giving all of the reviews equal weights and averaging makes a system vulnerable to spam and low-quality reviews. Therefore, PowerTrust selects representative and trusted reviewers and amplifies their influence during aggregation. We define trust level \(T\) in Eq.(21). Specifically, \(T\) is normalized to \([0,1]\) and computed from the buyer’s total number of transactions \(\rvert M \rvert\), average quality of the purchased data \(L\), and average review score \(R_i\) ($i=0,1...$), as calculated in Eqs.(19) and (20), respectively. Based on this score, we select the top \(1\%\) of buyers with the highest \(T\). Let \(M\) denote the set of transactions for each buyer, \(m \in M\) be a single transaction–review pair, and \(r_m\) be the review issued by the buyer for transaction \(m\). Using \(T\), we compute weight \(w_{{4i}}\) using Eq.(22). We compute the review score using the weight in Eq.(23) and normalize the result to obtain the final review value, \(R_4\). Let \(\mathcal{G}_i=\{\,H \mid H \text{ is a buyer who has evaluated provider } i\,\}\) be the set of buyers who rated provider \(i\), and let \(H\) index individual buyers. This approach mitigates the impact of low-quality reviews by emphasizing hub reviewers.

\begin{equation}
L = \frac{1}{\rvert M \rvert} \sum_{m \in M} \frac{1}{2}\bigl(q_\mathrm{acc} + q_\mathrm{comp}\bigr)
\end{equation}

\begin{equation}
R_i = \frac{1}{\rvert M \rvert} \sum_{m \in M} r_{m}
\end{equation}

\begin{equation}
T = \mathrm{normalize}\!\left( \frac{1}{3}(\rvert M \rvert + L + R_i) \right)
\end{equation}

\begin{equation}
w_{{4i}} = 0.5 + 0.5 T
\end{equation}

\begin{equation}
R_4 = \frac{\sum_{H \in \mathcal{G}_i} w_{{4i}} \, R_i}{\sum_{H \in \mathcal{G}_i} w_{{4i}}}
\end{equation}

\subsubsection{ PeerTrust}
PeerTrust integrates multiple factors (feedback values, reviewer reliability, transaction context, and time decay) into a single weighting scheme to compute the final reputation score of each provider. This method has been widely applied to opt-in markets with a low tamper resistance. For each completed transaction, the weighting uses the buyer’s review value, \(R_i\) (\(i=0,1,\ldots\)), and the quality of the traded data, \(L\), which is computed using Eq.(19), along with trust value \(T\) defined in Eq.(21) and Time-decay weight \(w_{{1i}}\) from Eq.(14). The final weight is determined using Eqs.(24) and (25). Using this weight, let \(\mathcal{J}_i=\{\,K \mid K \text{ is a buyer who evaluated provider } i\,\}\) be the set of buyers who rated provider \(i\), with \(K\) indexing individual buyers. The final review value \(R_5\) is computed using Eq.(26). This approach unifies the factors that were previously used separately, enabling the simultaneous consideration of transactional contextual consistency and recency.

\begin{equation}
I = \bigl(R_i \cdot T \cdot L\bigr)^{1/3}
\end{equation}

\begin{equation}
w_{{5i}} = w_{{1i}} \cdot I
\end{equation}

\begin{equation}
R_5 = \frac{\sum_{K \in \mathcal{J}_i} w_{{5i}} \, R_i}{\sum_{K \in \mathcal{J}_i} w_{{5i}}}
\end{equation}

\begin{table*}[t]
\centering
\caption{Parameters of Provider Strategies}
\begin{tabular}{l|cccc|c|l}
\hline
\multirow{2}{*}{\textbf{Strategy}} 
& \multicolumn{4}{c|}{\textbf{Quality Characteristics (mean $\pm$ std)}} 
& \multirow{2}{*}{\textbf{Price Mean}} 
& \multirow{2}{*}{\textbf{Behavioral Focus}}\\
\cline{2-5}
& \textbf{Accuracy} & \textbf{Compliance} & \textbf{Freshness} & \textbf{Coverage} &  &\\
\hline
\textbf{Trust} 
& 0.50 $\pm$ 0.10 & 0.60 $\pm$ 0.06 & 0.72 $\pm$ 0.10 & 0.70 $\pm$ 0.10 
& 72 $\pm$ 3
& Maintains moderate quality and low price to build trust.\\

\textbf{Price} 
& 0.70 $\pm$ 0.12 & 0.65 $\pm$ 0.10 & 0.70 $\pm$ 0.12 & 0.70 $\pm$ 0.12 
& 96 $\pm$ 4
& Prioritizes price competition; sacrifices quality for profit. \\

\textbf{Quality} 
& 0.80 $\pm$ 0.08 & 0.78 $\pm$ 0.08 & 0.78 $\pm$ 0.08 & 0.80 $\pm$ 0.08 
& 88 $\pm$ 4
& Pursues fair-value trades with high-quality data. \\

\textbf{Standard} 
& 0.75 $\pm$ 0.09 & 0.75 $\pm$ 0.09 & 0.75 $\pm$ 0.09 & 0.75 $\pm$ 0.09 
& 80 $\pm$ 3
& Balances all factors without specialization. \\
\hline
\end{tabular}
\label{tab:provider_strategies}
\end{table*}

\section{Experiment Setting}
\subsection{Simulation Process and Parameters}
The simulation modeled the interactions between $2,000$ providers and $2,000$ buyers in a single data marketplace with one platform operator (Fig. 1). Each simulation step represented one month, and the full run consisted of $120$ steps, corresponding to a $10$-year period of market activity. At the start of each run, provider reputations were initialized with values drawn from a normal distribution, and a reputation model was selected. In each step, providers and buyers selected strategies through Q-learning, conducted transactions, and updated their reputations and rewards. The simulation loop proceeded as follows.

\begin{table*}[t]
  \caption{Parameters of Simulation} 
  \label{tab:settings}
  \centering
  \captionsetup[subtable]{justification=centering,singlelinecheck=off}

  \begin{subtable}[t]{0.48\textwidth}
    \caption{Parameters of the Entire Simulation} 
    \label{tab:settings:a}
    \centering
    \renewcommand{\arraystretch}{1.15}
    \begin{tabular}{l|c|l}
      \hline
      \textbf{Parameter} & \textbf{Value} & \textbf{Description}\\
      \hline
      Providers/ Buyers     & 2,000 each & Market participants \\
      Marketplace/ Operator & 1 each     & Institutional agents \\
      Simulation steps      & 120        & Each step = 1 month \\
      Review values         & $[0,1]$    & Scaled feedback range \\
      Data price            & $>10$      & Means vary by provider strategy \\
      Data quality          & $[0,1]$    & Four-dimensional vector \\
      Fee / Rebate rate     & 10\%       & Exchange fee and rebate policy \\
      \hline
    \end{tabular}
  \end{subtable}
  \hfill
  \begin{subtable}[t]{0.48\textwidth}
    \caption{Parameters of Each Equation} 
    \label{tab:settings:b}
    \centering
    \renewcommand{\arraystretch}{1.15}
    \begin{tabular}{c|c!{\vrule width 1.5pt}c|c}
      \hline
      \textbf{Parameter} & \textbf{Value} & \textbf{Parameter} & \textbf{Value} \\
      \hline
      $b$ & $40$ & $o$ & $0.68$ \\
      $c$ & $0.2$& $u$ & $0.45$ \\
      $d$ & $0.6$& $v$ & $0.5$ \\
      $e$ & $0.6$& $y$ & $0.8$  \\
      $f$ & $0.2$& $\alpha$ & $0.15$  \\
      $g$ & $0.2$& $\beta$   & $0.92$ \\
      $h$ & $10$ & $\gamma$   & $0.9$ \\
      $k$ & $5$  & $\delta$ & $0.9$ \\
      $l$ & $0.31$  & $\epsilon$ & $1.00 \times 10^{-3}$  \\
          &         & $\zeta$ & $0.85$ \\
      \hline
    \end{tabular}
  \end{subtable}
\end{table*}

\begin{enumerate}
    \item Initialization: Reputation parameters and market attributes are set.
    \item Provider Action: Providers select one of four strategies and post data offers, determining price and quality (see Table III).
    \item Transaction and Review: Upon a successful trade, both parties pay a 10\% fee to the marketplace. Providers whose data meet high-compliance thresholds receive a 10\% fee rebate from the platform operator. Buyers then generate reviews based on transaction outcomes, updating the reputations of the providers according to the selected reputation system.
    \item Learning Update: Q-learning updates are applied for both providers and buyers using the new reward and state values, enabling continuous adaptation across time steps.
\end{enumerate}

This cycle was repeated until the simulation converged or was complete. The overall process is illustrated in Fig. 2. Tables II and III summarize the major simulation parameters. Unless otherwise noted, these values remained constant across all scenarios to ensure comparability.

\begin{table}[t]
\centering
\caption{Buyer Strategy Weights for Utility Calculation}
\setlength{\tabcolsep}{3pt}
\begin{tabular}{l|cccccc}
\hline
\multirow{2}{*}{\textbf{Strategy}} 
& \multicolumn{6}{c}{\textbf{Weight Parameters}} \\
\cline{2-7}
& Accuracy & Freshness & Coverage & Compliance & Reputation & Price \\
\hline
\textbf{Price} 
& 0.35 & 0.25 & 0.25 & 0.30 & 0.30 & 0.20\\

\textbf{Quality} 
& 0.15 & 0.15 & 0.15 & 0.20 & 0.20 & 0.45\\

\textbf{Trust} 
& 0.25 & 0.20 & 0.20 & 0.44 & 0.45 & 0.20\\
\hline
\end{tabular}
\label{tab:buyer_weights}
\end{table}

\begin{figure}[t]
    \centering
    \includegraphics[width=1.0\linewidth]{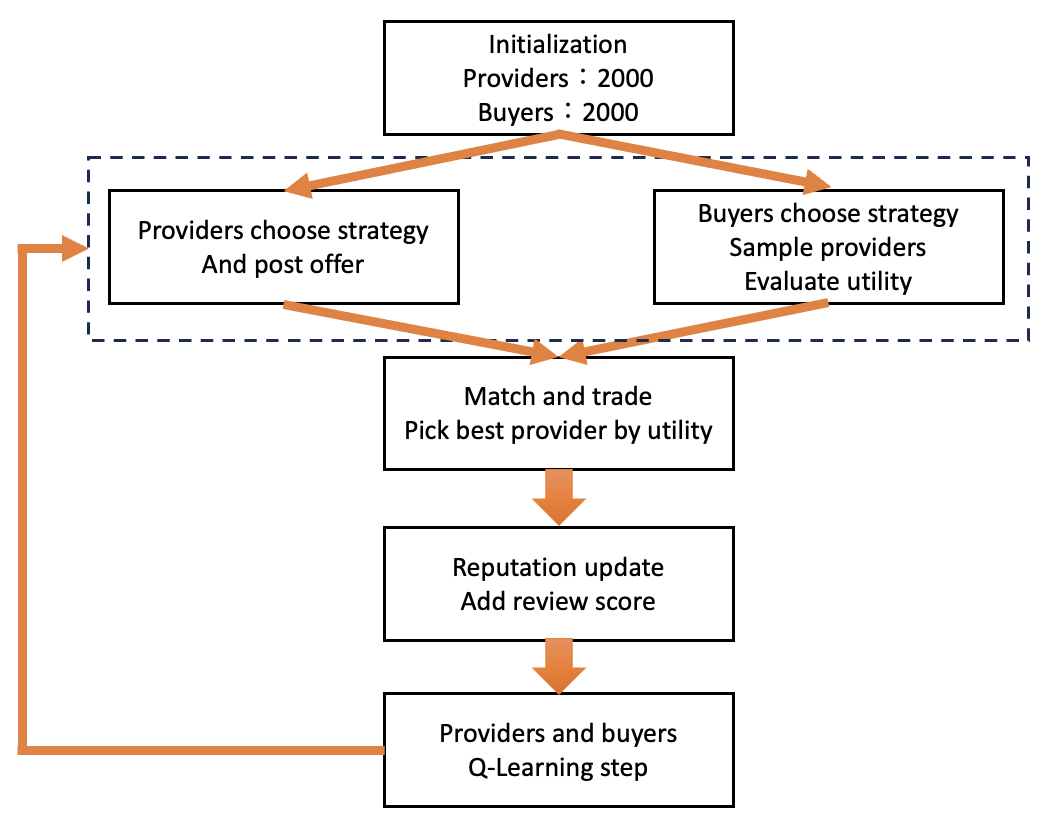}
    \caption{Simulation Process}
    \label{fig:placeholder}
\end{figure}

\subsection{Scenarios and Indicators}
We evaluated six scenarios to examine the market-level effects of reputation systems: Time-decay, Bayesian-beta, PageRank, PowerTrust, PeerTrust, and a no-reputation baseline (blind). Each scenario was simulated independently under identical market conditions. The outcomes were compared based on several macroscopic indicators.
\begin{itemize}
\item Welfare: Total revenue of the providers, platform operator, and marketplace
\item Avg Quality: Average quality of the data circulating in the market
\item Success Rate: Total number of transactions across all rounds, using the number obtained by multiplying the number of rounds by the total number of buyers as the denominator
\item Mean Price: Average price of all completed transactions
\item Platform revenue: Total exchange fee
\item Provider Sales Gini: Gini coefficient of provider revenue
\end{itemize}
The utility of the buyer was computed as the weighted sum of the quality attributes of the purchased dataset, prior-round reputation of the provider, and price. The six-dimensional weight vectors $[w_{\text{acc}}, w_{\text{fresh}}, w_{\text{cov}}, w_{\text{comp}}, w_{\text{rep}}$, $w_{\text{price}}]$ were used to determine the importance that buyers placed on each factor. These strategy weights differed with the strategy, as listed in Table~\ref{tab:buyer_weights}. In scenarios without a reputation system, $w_{\text{rep}}$ was omitted from the utility function, and the purchase threshold was adjusted downward to maintain consistent decision sensitivity.

\section{Results and Discussion}

\begin{figure*}[!t]
  \centering
  \begin{subfigure}[t]{0.48\textwidth}
    \centering
    \includegraphics[width=\linewidth]{\detokenize{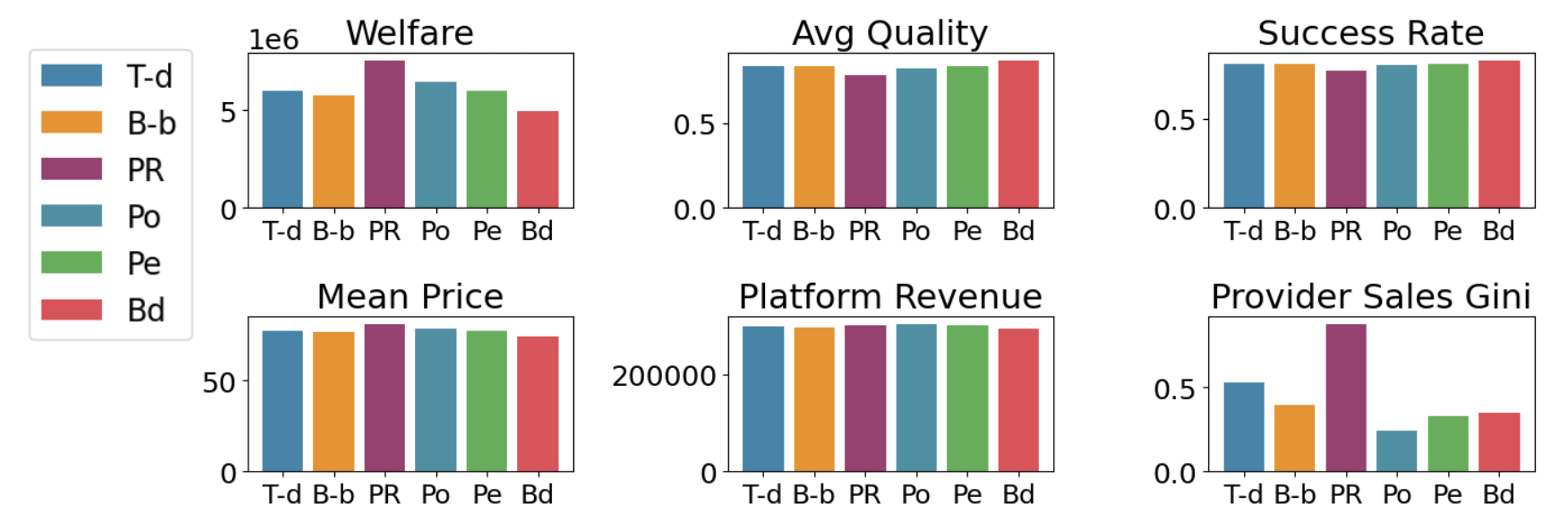}}
    \caption{Comparison of Market Indicators with and without Reputation Systems}
    \label{fig:indicators:a}
  \end{subfigure}%
  \hfill
  \begin{subfigure}[t]{0.48\textwidth}
    \centering
    \includegraphics[width=\linewidth]{\detokenize{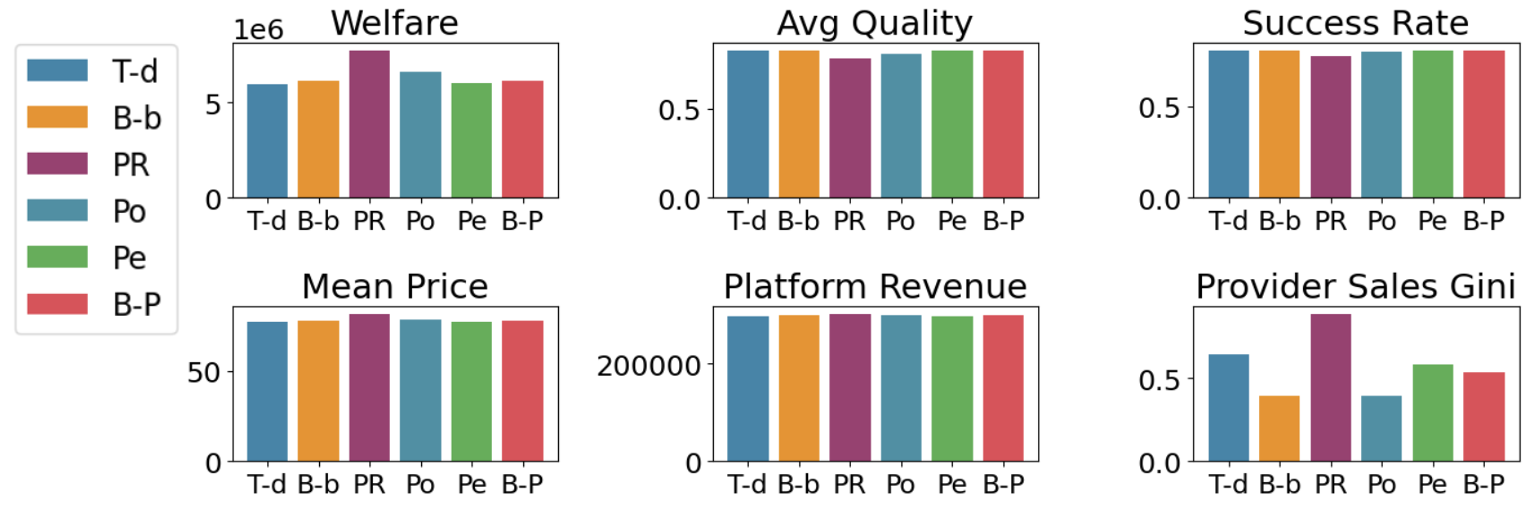}}
    \caption{Comparison of Existing Methods and New Method}
    \label{fig:indicators:b}
  \end{subfigure}
  \caption{Market indicators}
  \label{fig:indicators}

  \centering
  \begin{subfigure}[t]{0.48\textwidth}
    \centering
    \includegraphics[width=\linewidth]{\detokenize{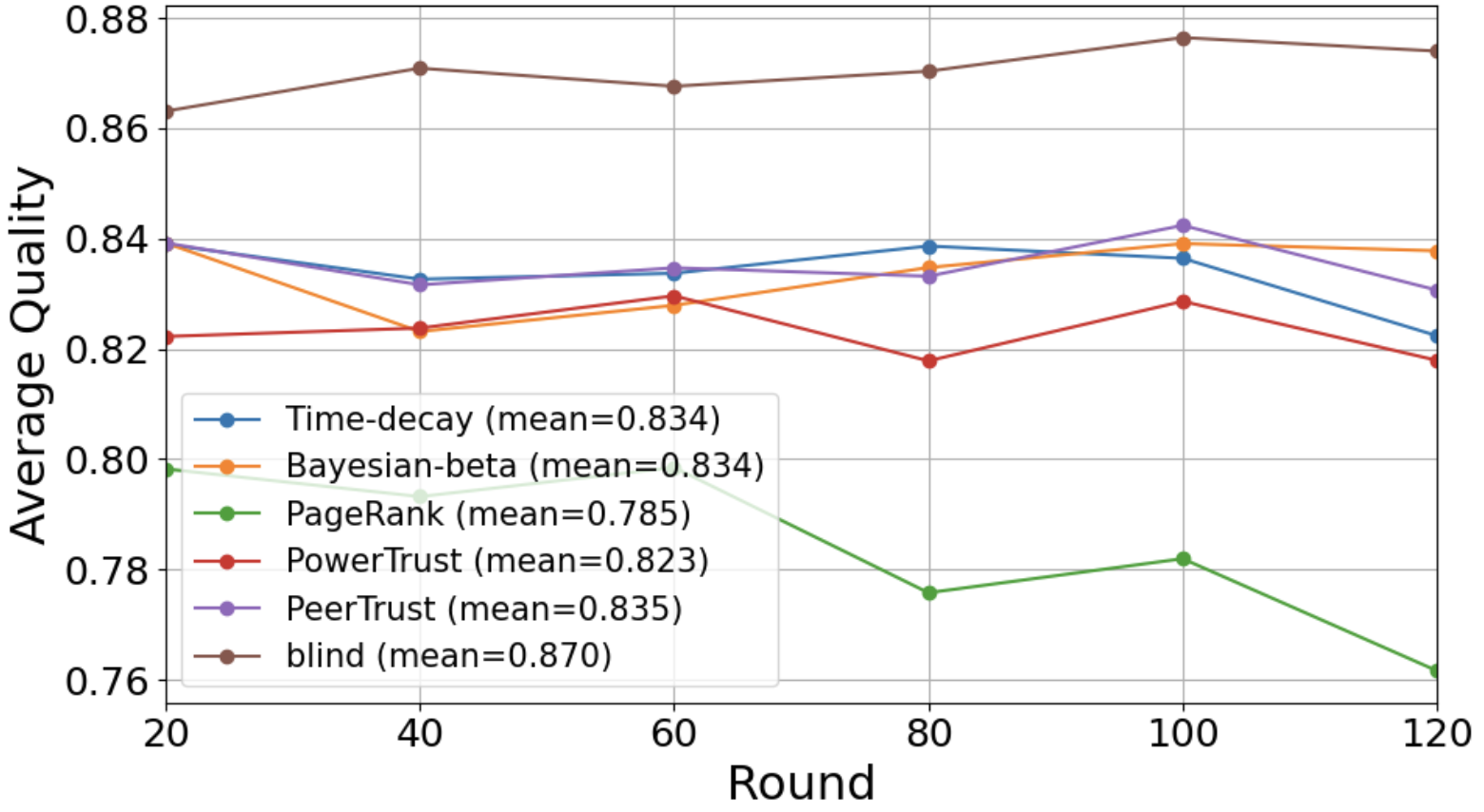}}
    \caption{Comparison of Markets with and without Reputation Systems}
    \label{fig:indicators:a}
  \end{subfigure}%
  \hfill
  \begin{subfigure}[t]{0.48\textwidth}
    \centering
    \includegraphics[width=\linewidth]{\detokenize{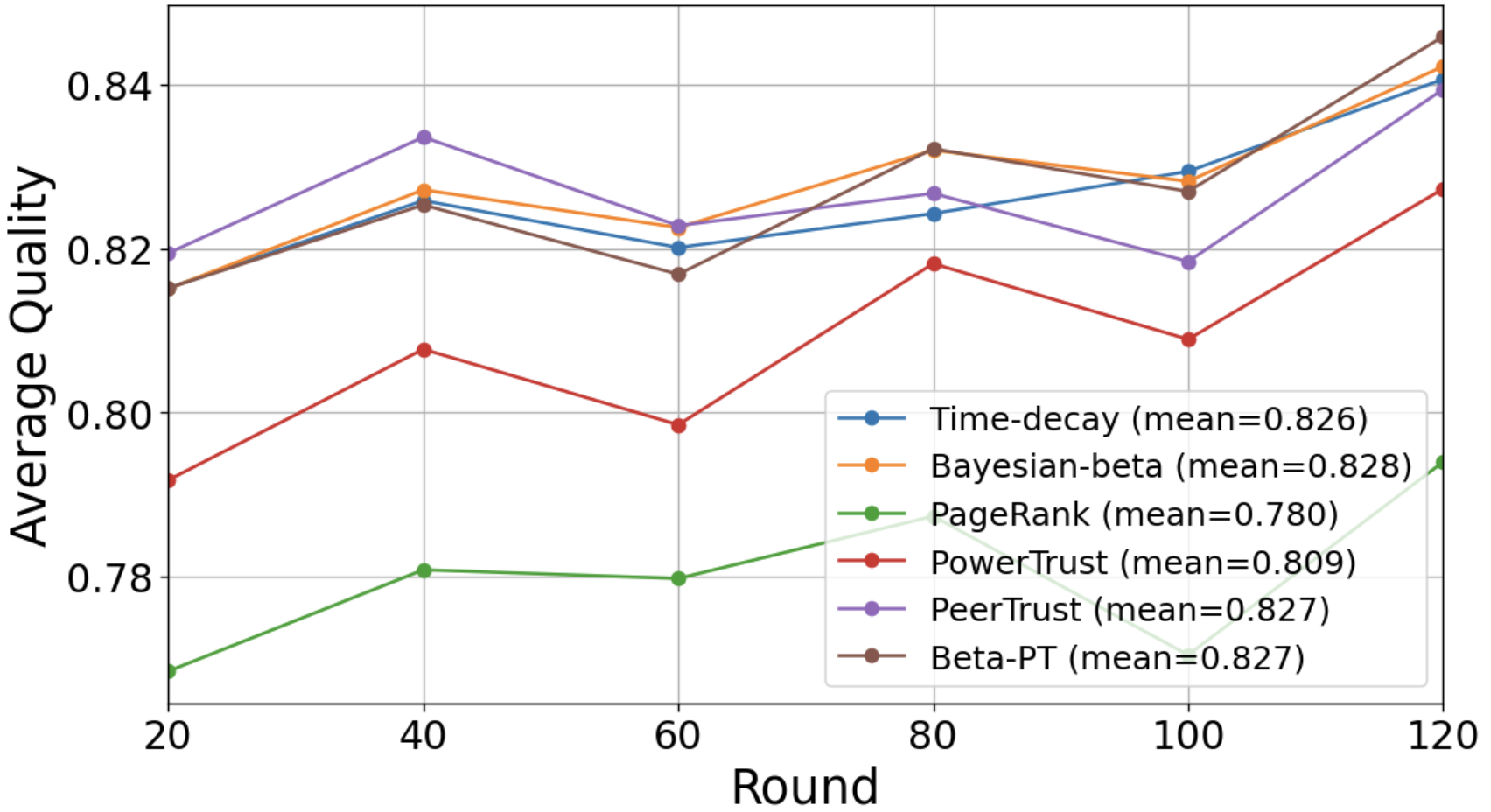}}
    \caption{Comparison of Existing Methods and New Method}
    \label{fig:indicators:b}
  \end{subfigure}
  \caption{Average Trends of Data Quality Values}
  \label{fig:indicators}

  \centering
  \begin{subfigure}[t]{0.48\textwidth}
    \centering
    \includegraphics[width=\linewidth]{\detokenize{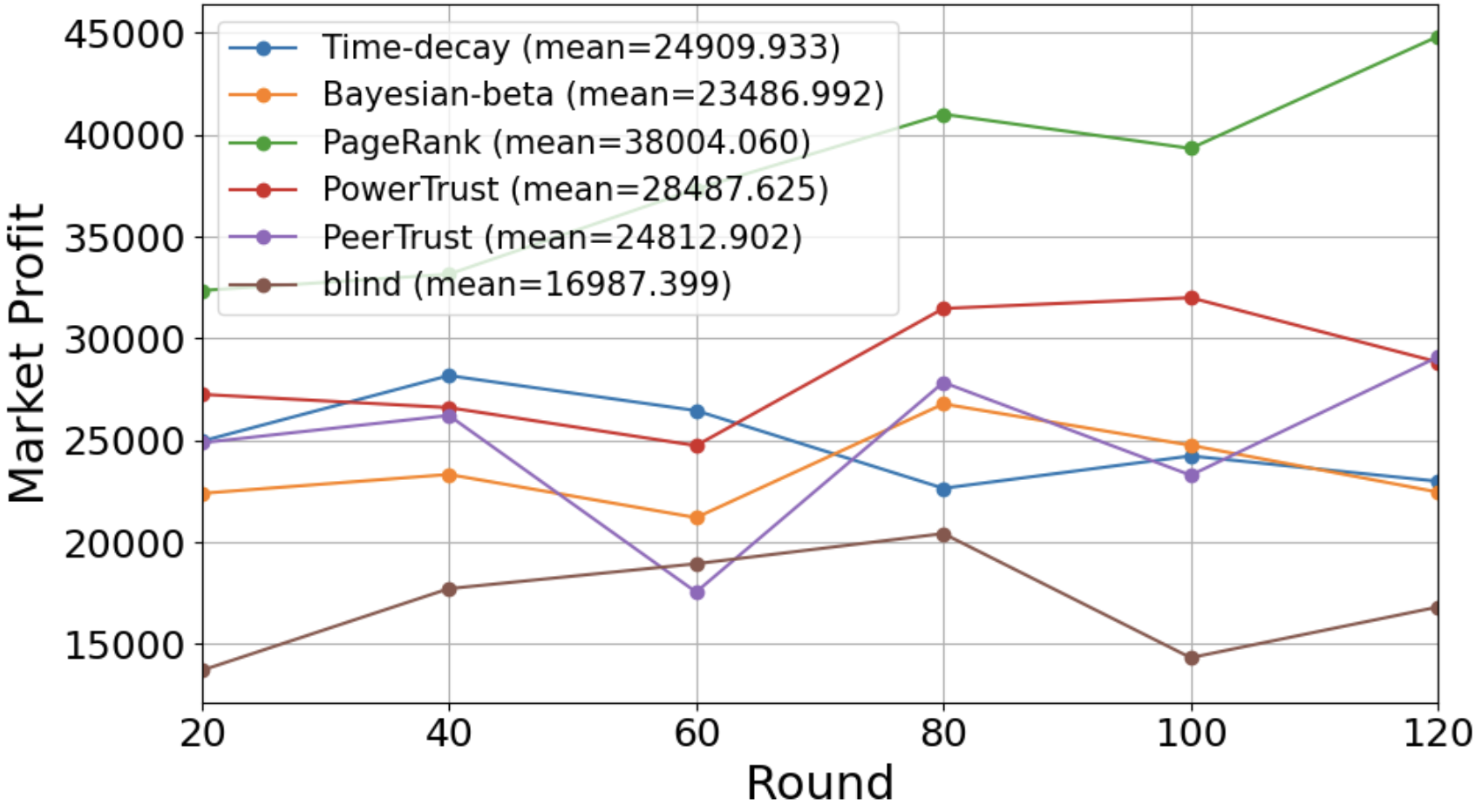}}
    \caption{Comparison of Markets with and without Reputation Systems}
    \label{fig:indicators:a}
  \end{subfigure}%
  \hfill
  \begin{subfigure}[t]{0.48\textwidth}
    \centering
    \includegraphics[width=\linewidth]{\detokenize{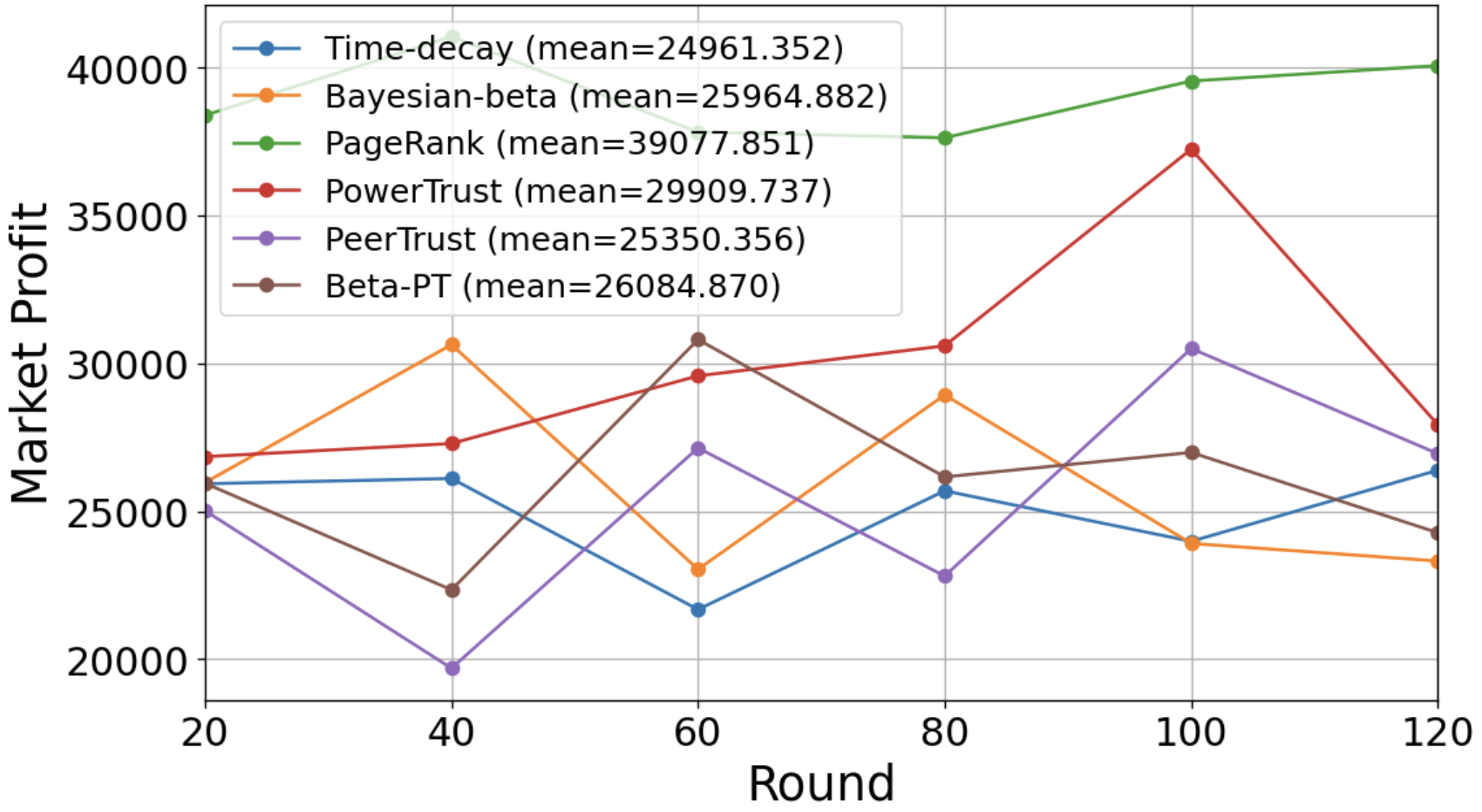}}
    \caption{Comparison of Existing Methods and New Method}
    \label{fig:indicators:b}
  \end{subfigure}
  \caption{Average Trends of Market Profitability}
  \label{fig:indicators}

    \centering
  \begin{subfigure}[t]{0.48\textwidth}
    \centering
    \includegraphics[width=\linewidth]{\detokenize{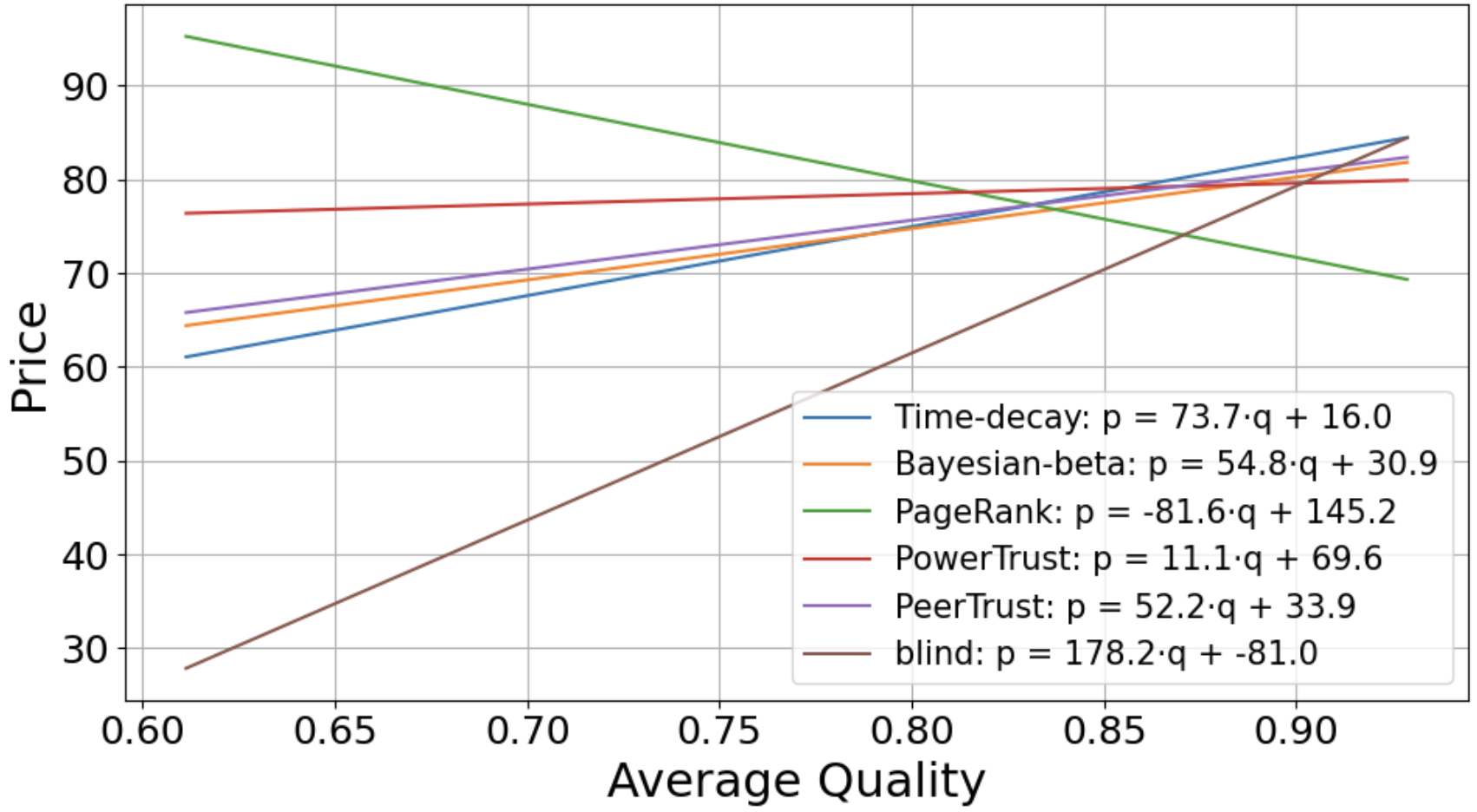}}
    \caption{Comparison of Markets with and without Reputation Systems}
    \label{fig:indicators:a}
  \end{subfigure}%
  \hfill
  \begin{subfigure}[t]{0.48\textwidth}
    \centering
    \includegraphics[width=\linewidth]{\detokenize{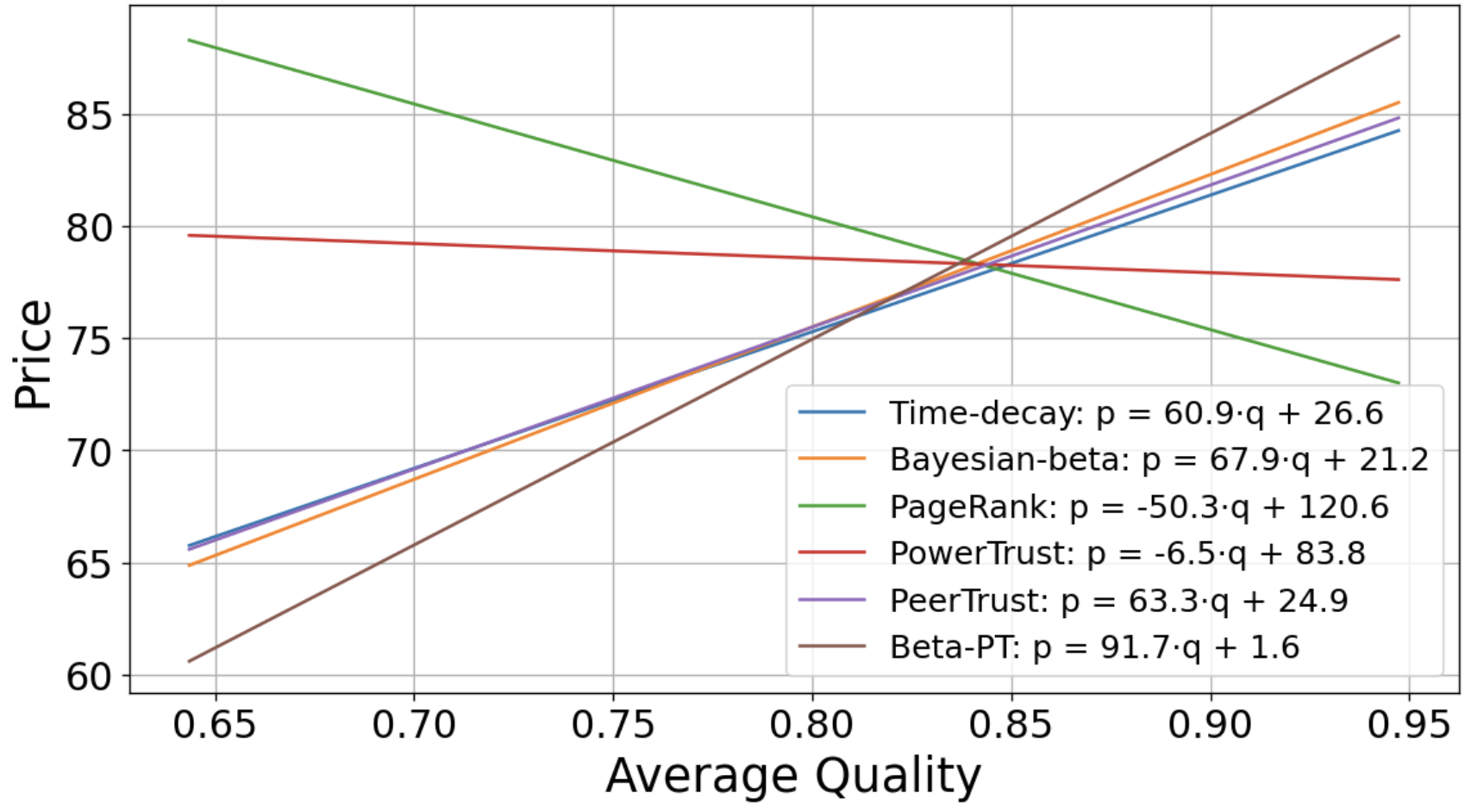}}
    \caption{Comparison of Existing Methods and New Method}
    \label{fig:indicators:b}
  \end{subfigure}
  \caption{Price-Quality Frontiers}
  \label{fig:indicators}
\end{figure*}

\subsection{Six Scenarios Comparing Five Reputation Systems}
Figure 3 (a) compares the market performances across all scenarios. Welfare—the total profits of the providers, marketplace, and platform operator, was the highest in the PageRank scenario. This result indicated that a small number of highly ranked providers attracted disproportionate transaction volumes, generating high overall profits. However, the Gini coefficient, which measured the inequality of provider revenues, was also the highest under PageRank, indicating a strong market concentration. By contrast, PowerTrust and PeerTrust exhibited notably lower Gini values, which suggested a more balanced distribution of transactions among providers. The baseline no-reputation market (blind) exhibited the lowest overall welfare, highlighting the importance of trust mechanisms in stimulating economic activities.

Figure 4 (a) illustrates the evolution of the average data quality circulation in the market over time. To make it easier to observe the changes, the results were plotted as the averages of every $20$ steps. Although the no-reputation market had the highest average quality, this was because buyers base their decisions solely on observable quality and prices in the absence of trust information. Conversely, the PageRank market had the lowest overall data quality, because transactions were heavily skewed toward a few dominant providers whose data did not necessarily maintain consistent quality levels. By contrast, markets using Time-decay, Bayesian-beta, or PeerTrust maintained moderate yet stable levels of data quality throughout the simulation, suggesting that the reputation system effectively sustained trust and prevented severe quality degradation.

Figure 5 (a) plots the temporal changes in provider revenue. To make it easier to observe the changes, the results were plotted as averages taken for every $20$ steps, just as in Fig. 4. Although fluctuations occurred in every scenario, the PageRank system yielded the highest average revenue, but also exhibited significant volatility, which was consistent with its high market concentration. By contrast, PeerTrust and Bayesian-beta showed more stable revenue trajectories, implying that the inclusion of the transaction context and reviewer credibility contributed to greater market equilibrium. By comparison, the no-reputation market produced the lowest and most erratic revenue levels, reflecting the instability of unregulated exchanges.

Figure 6 (a) shows the relationship between the data quality (x-axis) and price (y-axis) based on the results of a linear regression analysis across all reputation systems. Markets with a high alignment between quality and price are more likely to be sustainable and stable [38]. The results indicated that the slope was the steepest in market without a reputation, suggesting that buyers place excessive emphasis on observable quality when reliable information is lacking. By contrast, PageRank exhibited the smallest slope, meaning that provider dominance decouples price from quality. Among the models, Time-decay, Bayesian-beta, and PeerTrust showed relatively large slopes, reflecting well-functioning markets with high consistency between price and quality.

\subsection{Proposal of a New Reputation System}
Building on the market-simulation results discussed in the previous subsection, we propose a new hybrid reputation system tailored to the characteristics of data-trading markets. The preceding experiments revealed that Time-decay, Bayesian-beta, and PeerTrust achieved the most stable market performances. However, the consistency between the data price and quality remained lower than that in the no-reputation baseline.

Among these, PeerTrust was found to balance trust propagation and market stability most effectively because it integrated multiple evaluation factors, including reviewer credibility, transaction context, and the data quality of recent trades. Conversely, Bayesian-beta mitigated rating inflation by suppressing extremely early reviews and overactive reviewers. By combining these complementary properties, we designed a hybrid system with the goal of achieving both stability and price–quality alignment.

Accordingly, we propose a new model called Beta-PT, which extends the PeerTrust framework by incorporating two mechanisms inspired by Bayesian-beta: (i) the exclusion of excessive weighting on recent individual reviews and (ii) introduction of mild penalties for newcomers, while damping the influence of overactive reviewers. Formally, we define review confidence as follows:

\begin{equation}
\mathrm{conf} = \frac{A + D}{A + D + 4},
\end{equation}

\noindent where $A$ and $D$ denote the numbers of positive and negative evaluations, respectively. This confidence factor moderates the volatility of reputation updates, particularly for providers with limited transaction histories. The modified weight for each transaction was computed as follows:

\begin{equation}
w_{{6}} = w_{{1i}} \cdot T \cdot L \cdot \mathrm{conf},
\end{equation}

\noindent where $w_{{1i}}$ represents the time-decay weight, $T$ is the trust factor in Eq.(21), and $L$ is the transaction quality in Eq.(19). To prevent the dominance of a small subset of frequent reviewers, the following corrective weight was applied:

\begin{equation}
w_z=
\begin{cases}
1, & \text{if } W=0 \text{ or } w_{{6}}=0\\[4pt]
\min\!\Bigl(1,\ \dfrac{z\,W}{w_6}\Bigr), & \text{otherwise},
\end{cases}
\end{equation}

\noindent where $W$ denotes the total number of reviews for a provider, $w_z$ is the number submitted by a particular buyer, and $z = 0.35$ is a damping coefficient. This formulation effectively curbs review monopolization while maintaining a balance for the influence of feedback sources.

The proposed Beta-PT system was evaluated under the experimental conditions described in Section IV. Figures 3 (b) – 6 (b) present four outcome indicators: the welfare, data quality, market profitability, and price–quality correlation.

The results demonstrated that Beta-PT achieved lower provider dominance than PeerTrust, indicating reduced market concentration. In addition, the market profitability and data quality were comparable to those of the best existing systems (Figs.  4 (b) and 5 (b)). Moreover, the highest price–quality consistency among all methods (Fig. 6 (b)) indicated that prices  reflected the data values more accurately.

Collectively, these results suggested that the new system has the potential to foster a market in which high-quality data are traded at appropriate premium prices, thereby cultivating trust and economic efficiency. Furthermore, mitigating the excessive influence of a small number of high-frequency reviewers will support fairer evaluation dynamics and more equitable market participation. However, because this study did not examine the rationale or validity of the parameters used in the simulations, or conduct sensitivity analyses, verifying these aspects will be a future task to strengthen the claims of this paper. 

\section{Conclusion}
This study examined the institutional design of data-trading markets by focusing on how reputation mechanisms influence trust, quality, and market stability. We developed a multi-agent simulator that integrated RL and IRL to model realistic, and evaluated five existing reputation systems: Time-decay, Bayesian-beta, PageRank, PowerTrust, and PeerTrust. Based on the results, we proposed a new hybrid reputation model (Beta-PT) that combined the contextual evaluation of PeerTrust with the credibility weighting of Bayesian-beta.

The results showed that, while the introduction of any reputation system slightly lowers the average data quality, it increases market profitability by enabling providers to monetize their reputation rather than competing solely based on prices.
Among the models, Beta-PT achieved the highest price–quality consistency and reduced market concentration, fostering markets in which high-quality data are traded at fair prices.

These findings suggested that multi-factor adaptive trust mechanisms are essential for sustaining efficient and equitable data-trading markets.
Future research should extend this framework by incorporating more diverse agent strategies, demand-responsive pricing, and the behaviors of non-reviewing or adversarial participants.
Overall, this study provided methodological and institutional insights into how trust-based reputation systems can enhance the reliability and efficiency of emerging data-trading ecosystems.

\section*{Acknowledgment}
This study was supported by the JST PRESTO Grant Number JPMJPR2369 and JSPS KAKENHI Grant Number JP25K00153.

\end{document}